\documentclass[conference]{IEEEtran}
\IEEEoverridecommandlockouts
\usepackage{cite}
\usepackage{amsmath,amssymb,amsfonts}
\usepackage{algorithmic}
\usepackage{graphicx}
\usepackage{textcomp}
\usepackage{xcolor}
\def\BibTeX{{\rm B\kern-.05em{\sc i\kern-.025em b}\kern-.08em
    T\kern-.1667em\lower.7ex\hbox{E}\kern-.125emX}}

\usepackage{epstopdf}
\renewcommand\phi{\varphi}
 \DeclareMathOperator{\col}{col}
 
\newcommand{\trn}{^{\rm\scriptscriptstyle T}}

\DeclareMathOperator{\e}{e}

\begin{document}

\title{Synchronization in Networks of Neural Mass Model Populations with Discrete Couplings
\thanks{The proof of Theorem~\eqref{th2} was performed in IPME RAS and supported by the Ministry of Science and
Higher Education of the Russian Federation (Project No. 075-15-2021-573).

The simulation (SubSec~\ref{sec:sim}) was performed in Lobachevsky State University of Nizhny Novgorod and supported by the Russian Science
Foundation (Project No. 19-72-10128).}
}

\author{\IEEEauthorblockN{Sergei A. Plotnikov}
\IEEEauthorblockA{\textit{Institute for Problems of Mechanical Engineering, RAS}\\
Saint-Petersburg, Russia \\
\textit{Lobachevsky State University of Nizhny Novgorod}\\
Nizhny Novgorod, Russia \\
e-mail: waterwalf@gmail.com}
}

\maketitle

\begin{abstract}
The problem of synchronization in networks of neural mass model populations with discrete couplings is considered. The considered network is hybrid one, therefore Mikheev approach is applied to transform it to the network with time-varying delayed couplings. Thus the problem of hybrid network synchronization is reduced to the studying of synchronization in networks with delayed couplings, which was previously solved by analytical means. It is showed that the Laplace matrix spectrum and maximum sampling interval are defining for networks dynamics. The dynamics of $5$ neural mass model populations with discrete couplings was simulated for $3$ different situations. The first case deal with the asymptotic synchronization, when both maximum eigenvalue of Laplacian and maximum sampling interval are small enough. The second case is about $\varepsilon$-synchronization, which is achieved for small enough maximum eigenvalue of Laplacian and big sampling intervals. And the last case is desynchronization of oscillations, which has been observed for big values of Laplacian eigenvalues and sampling intervals.
\end{abstract}


\section{Introduction}
To study the dynamics of different systems they often supposed to be smooth and continuous. On the other hand, variables such as friction, switching and sliding can cause intricate response and abrupt changes. To account for these features the nonlinear systems with the coexistence of discrete and continuous dynamics are considered. This type of system has a technical term "Hybrid system" \cite{MAT00,LUN09}. The hybrid systems has rich dynamical nature with respect to a smooth continuous systems \cite{KOB07,HAD06}. In particular, a hybrid modeling handle the switching properties in biochemical systems and gene regulatory networks \cite{PER10,AIH10}. Neurons interact with each other by
spiking, which occurs between the resting periods \cite{PLO16b}.

One of the most attractive phenomenon in network dynamics is synchronization. An area of special interest is synchronization in large populations of interacting oscillatory elements \cite{KUR84,TAS99,STR00}. Examples of synchronization, include, among others, numerous forms of collective behavior in complex biological and technical formations, such as a swarm of insects, a flock of birds \cite{HER16,SUM10}; ensembles of coupled oscillators \cite{HON11} and a group of mobile robots \cite{REN08}. Special attention is paid to synchronization in neural network dynamics, where it is connected with brain cognitive abilities \cite{MUR96,SIN00} and pathological brain states \cite{HAM07,MIL03}.

In this paper we continue the work started in \cite{PLO21, PLO22}. In these works the synchronization problem of linear systems with nonlinear non-delayed and delayed couplings was considered. As an example the dynamics of neural mass model (NMM) population network \cite{JAN95} was investigated. This model describes the mean activity of entire neural populations, represented by their averaged firing rates and membrane potentials. It was designed specifically for modeling of electroencephalography rhythms and evoked potentials \cite{KRO09}. The approach to study the synchronization problem in networks of such systems is based on coordinate transformation proposed in \cite{PAN17} and application of circle criterion and its time-delay version \cite{CHU95,BRY19} to the synchronization error system stability analysis. In this paper we focus on the case of the network of NMMs with discrete couplings. Note that in \cite{PLO16b} the synchronization problem of other neuron models, namely FitzHugh-Nagumo model, with discrete couplings was studied.

The rest of the paper is organized as follows. Section~\ref{sec:prel} gives a brief information about NMM system, graph theory and synchronization of delay coupled NMMs. Subsection~\ref{sec:main} shows how to present hybrid NMM network as continuous system with time-varying delay and obtain its synchronization conditions. In Subsec~\ref{sec:sim} numerical results on asymptotic synchronization, $\varepsilon$-synchronization and desynchronization are provided. Finally, conclusions are given in Sec.~\ref{sec:conclusion}.

\section{Preliminaries}\label{sec:prel}
\subsection{Model Equation}

The NMM is introduced by Jansen and Rit in 1995 \cite{JAN93,JAN95} and is used for electrical brain activity simulation using macroscopic parameters such as the average membrane potential and firing rate. This model is originally used for simulation of alpha rhythm and evoked potentials and further was improved to produce richer rhythms and to study epileptic activity \cite{DAV03}. NMM models the average firing activity of a pyramidal neuron population that interacts with two populations of interneurons and combines inhibitory and excitatory signals from them \cite{JAN95}. In this paper the NMM with discrete coupling is considered, i.e. we deal with a hybrid system. The dynamics of each neuronal population is described by the second-order nonlinear hybrid equation in the following form:
$$
\ddot y(t) +2\alpha \dot y(t) +\alpha ^2y(t) - \alpha\beta\phi(\sigma(t_k))=0, 
$$
where $y$ is an output of the system (the post-synaptic potential); $\sigma(t_k)=\col\{\sigma_1(t_{1}^k),\dots,\sigma_m(t_{m}^k)\}$ is a discrete input, while $t_{i}^k$ are discrete moments of time, $i=1,\dots,m$, \mbox{$k=0,1,2,\dots$}. $\alpha$ is the reciprocal of the synaptic time constant; $\beta$ is the gain for the post-synaptic response kernel. The values of parameters $\alpha$ and $\beta$ define the excitatory or inhibitory behavior of the population. As $\phi(\sigma)$ a nonlinear centered sigmoidal function is considered:
$$
\phi(\sigma) = \frac{2e_0}{1+\e^{r(v_0-\sigma)}}-\frac{2e_0}{1+\e^{rv_0}},
$$
where $e_0>0$ represents the maximum firing rate of the population, $v_0$ is the ratio of average inhibitory synaptic gain and $r$ is the steepness of sigmoidal function. Here we choose the standard values of $e_0=2.5$ and $r=0.56$ \cite{JAN95}, and  $v_0=0$. In this case the sigmoid graph lies in the sector between two straight lines $0$ and $0.5e_0r\sigma$ (see explanation in \cite{GOR17}).

Denoting $x=\dot y$ the NMM can be presented in the state space:
$$
\begin{aligned}
    \dot y(t) &=x(t), \\
    \dot x(t) &=-2\alpha x(t) -\alpha ^2y(t) + \alpha\beta\phi(\sigma(t_k))=0.
\end{aligned}
$$

\subsection{Graph Theory}

Graph theory is usually used to describe the structural properties of the network. A directed graph is an ordered pair $G = (V, E)$, where $V$ corresponds to a set of $N$ nodes, while $E \subseteq V \times V$ denotes a set of edges. The graph is undirected if for each edge $(v_1,v_2)\in E$, where $v_1, v_2\in V$, $(v_2,v_1)\in E$ is fulfilled. A node $v_1$ is connected to a node $v_2$ in the graph $G$ if $(v_1, v_2) \in E$. A path in the graph is a finite sequence of nodes $v_1,\dots, v_k$, if any element of the sequence is connected with the following one $(v_i,v_{i+1})\in E$ for $i<k$). The undirected graph $G$ is connected, if it contains a path from $v_1$ to $v_2$ for all $(v_1,v_2)\in E$. A weighted graph is a graph in which each edge $(v_i,v_j)$ has its numerical weight $a_{ij}$ such that:
$$
\begin{cases}a_{ij}>0, \mbox{ if } (v_i,v_j)\in E,\\
a_{ij}=0, \mbox{ if } (v_i,v_j)\notin E.
\end{cases}
$$
These weights form the adjacency matrix $A=[a_{ij}]$. Below, everywhere the graphs without loops are considered, i.e. $(v,v)\notin E$ for all $v\in V$. The Laplace matrix of graph $G$ has the following form
$$
L=\begin{bmatrix} 
\sum_{j=2}^{N}a_{1j} &-a_{12} &\cdots&-a_{1N}\\
-a_{21} &\sum_{j=1,j\ne2}^{N}a_{2j}&\cdots&-a_{2N} \\
\vdots &\vdots &\ddots&\vdots \\
-a_{N1} &-a_{N2} &\cdots&\sum_{j=1}^{N-1}a_{Nj}
\end{bmatrix}.
$$
This matrix is symmetric $L=L\trn$ for undirected graph $G$. Also it always has zero minimal eigenvalue $\lambda_1(G)=0$, in particular, $L(G)\ge0$ \cite{AGA09,OLF07} (here and below, the notation $L\ge0$ for a symmetric matrix $L$ denotes its non-negative definiteness). The second minimal eigenvalue of Laplacian $\lambda_2(G)$ is called algebraic connectivity of graph $G$ \cite{FIE73}. $\lambda_2(G)>0$ if and only if undirected graph $G$ is connected.
The properties of Laplace matrices plays a crucial role in dynamics of network with diffusive couplings, since diffusive coupling can be described using Laplace matrix.

\subsection{Synchronization in Delay Coupled Neural Mass Model Populations}\label{sec:synch}
Consider the network of delay coupled NMM populations:
\begin{equation} \label{net}
\begin{aligned}
\dot y_{i}(t) &= x_{i}(t), \\
\dot x_{i}(t) &= -2\alpha x_{i}(t) -\alpha ^2y_{i}(t)\\
&+\alpha\beta\phi\left[\sum\limits_{j=1}^N a_{ij}\left(x_{i}(t-\tau_{ij}(t))-x_{j}(t-\tau_{ij}(t))\right)\right]. 
\end{aligned}
\end{equation}
Here $z_i=\col\{x_i,y_i\}$ is a state vector of $i$th node; $a_{ik}$ are coupling coefficients, which form the adjacency matrix \mbox{$A=[a_{ij}]$}; $\tau_{ij}(t)\in[0,T]$, $\forall t$, $i,j=1,\dots,N$ are bounded time-varying delays (all delay functions have the same upper bound $T$). This type of coupling is diffusive one, and both signals are delayed. For example, a centralized control law has such type of coupling.

This paper studies a synchronization problem, which is the asymptotically identical evolution of the systems \cite{BLE88}. Here we consider a coordinate synchronization \cite{FRA07}:
\begin{equation}\label{goal1}
\lim\limits_{t\to\infty}(z_i(t)-z_j(t))=0,\quad i,j=1,\dots,N.
\end{equation}
Achieving the goal \eqref{goal1} leads to the same behavior of each node \eqref{net}, which can be described by mean-field dynamics $z_s$. Verification of the condition \eqref{goal1} is quite complicated with respect to the analytical analysis. Therefore the problems of network \eqref{net} synchronization can be reduced to studying stability of systems' synchronization errors $e_i=z_i-z_s$:
\begin{equation}\label{goal2}
\lim\limits_{t\to\infty}e_i(t)=0,\quad i=1,\dots,N.
\end{equation}

The problem of synchronization of non-linearly delay coupled linear systems was considered in \cite{PLO22}, in particular, synchronization conditions of delay coupled NMM populations were derived. The approach is based on circle criterion for time delay systems \cite{CHU95,BRY19}, which is the generalization of approach given in \cite{PLO21} for time delay case.

\newtheorem{thm}{Theorem}
\begin{thm}\label{th1}
If the network \eqref{net} systems parameters $\alpha>0$ and $\beta>0$, the graph of the network is connected and undirected, and the following inequalities are fulfilled:
$$
    \lambda_{\max}<\frac{1}{\beta re_0},\quad T<\frac{2}{re_0\lambda_{\max}\alpha\beta},
$$
where $\lambda_{\max}$ is the maximum eigenvalue of the Laplace matrix $L$. Then the network of delay coupled NMM populations \eqref{net} is asymptotically synchronized, i.e. the goal \eqref{goal2} is fulfilled.
\end{thm}

\section{Main Result}
\subsection{Synchronization in Hybrid Neural Mass Model Populations}\label{sec:main}
Consider the network of $N$ NMM populations with discrete couplings:
\begin{equation} \label{netm}
\begin{aligned}
\dot y_{i}(t) &= x_{i}(t), \\
\dot x_{i}(t) &= -2\alpha x_{i}(t) -\alpha ^2y_{i}(t)\\
&+\alpha\beta\phi\left[\sum\limits_{j=1}^N a_{ij}\left(x_{i}(t_{ij}^k)-x_{j}(t_{ij}^k)\right)\right]. 
\end{aligned}
\end{equation}
It is assumed that $t_{ij}^{k+1}-t_{ij}^k= h_{ij}$, $i,j=1,\dots,N$, where $h_{ij}>0$ are sampling intervals for various pairs of nodes $(i,j)$. Suppose that the graph of presented network is connected and undirected, i.e. its adjacency matrix is symmetric ($a_{ij}=a_{ji}$, $i\ne j$, $i,j=1,\dots,N$). If the sampling interval between the node $i$ and $j$ is equal to $h_{ij}$, then the sampling interval between the node $j$ and $i$ has the same value $h_{ij}$.

Following the Mikheev approach \cite{MIK88,FRI92} the system \eqref{netm} can be presented as a continuous system with time-varying delays $\tau_{ij}(t)=t-t_{ij}^k$ \eqref{net}, where the delays are piecewise-linear (sawtooth) with $\dot\tau_{ij}(t)=1$ for $t\ne t_{ij}^k$. Therefore \mbox{$0\le \tau_{ij}(t)\le h_{ij}$}, $i,j=1,\dots,N$, and there exist the maximum upper bound for all delays $T=\max_{i,j=1,\dots,N}h_{ij}$. Modeling of continuous-time systems with discrete blocks in the form of continuous-time systems with time-varying delay have allowed to apply the time-delay approach to hybrid systems. Thus, the results of the Theorem~\ref{th1} are valid for transformed hybrid network \eqref{netm}. The following theorem holds.

\begin{thm}\label{th2}
If the network \eqref{netm} systems parameters $\alpha>0$ and $\beta>0$, the graph of the network is connected and undirected, and the following inequalities are fulfilled:
\begin{subequations}
\begin{align}\label{res1}
\lambda_{\max}&<\frac{1}{\beta re_0}, \\
h&<\frac{2}{re_0\lambda_{\max}\alpha\beta}\label{res2}
\end{align}
\end{subequations}
where $\lambda_{\max}$ is the maximum eigenvalue of the Laplace matrix $L$ and $h=\max_{i,j=1,\dots,N}h_{ij}$ is the maximum sampling interval. Then the network of hybrid NMM populations \eqref{netm} is asymptotically synchronized, i.e. the goal \eqref{goal2} is fulfilled.
\end{thm}

\subsection{Simulation}\label{sec:sim}

\begin{figure}
\center{\includegraphics[width=0.8\linewidth]{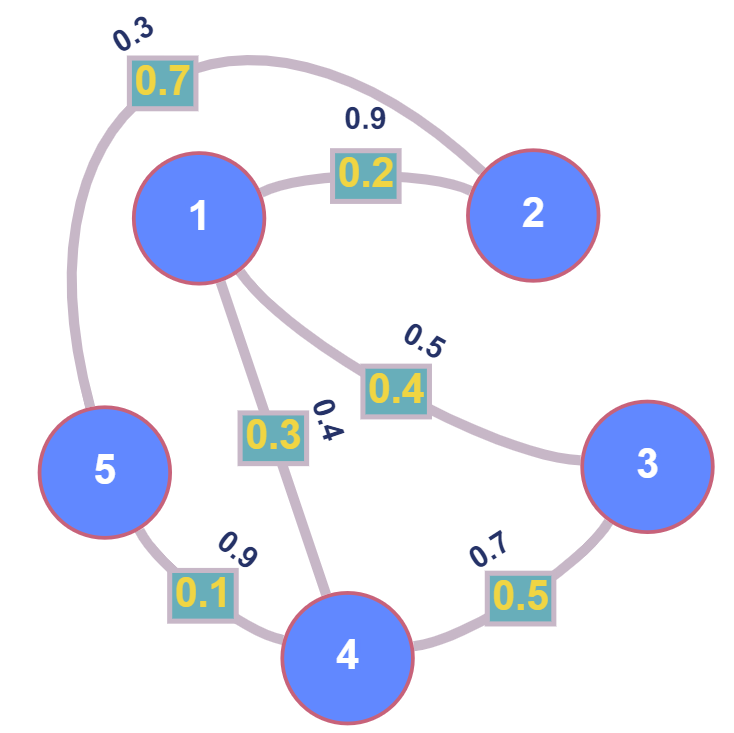}}
\caption{Undirected graph with $N=5$ vertices.}
\label{fig1}
\end{figure}
This section presents the results of simulation. The simulation was performed in MATLAB \cite{MAT} using the Hybrid Equations (HyEQ) Toolbox \cite{HYE}.

\begin{figure}
\center{\includegraphics[width=1\linewidth]{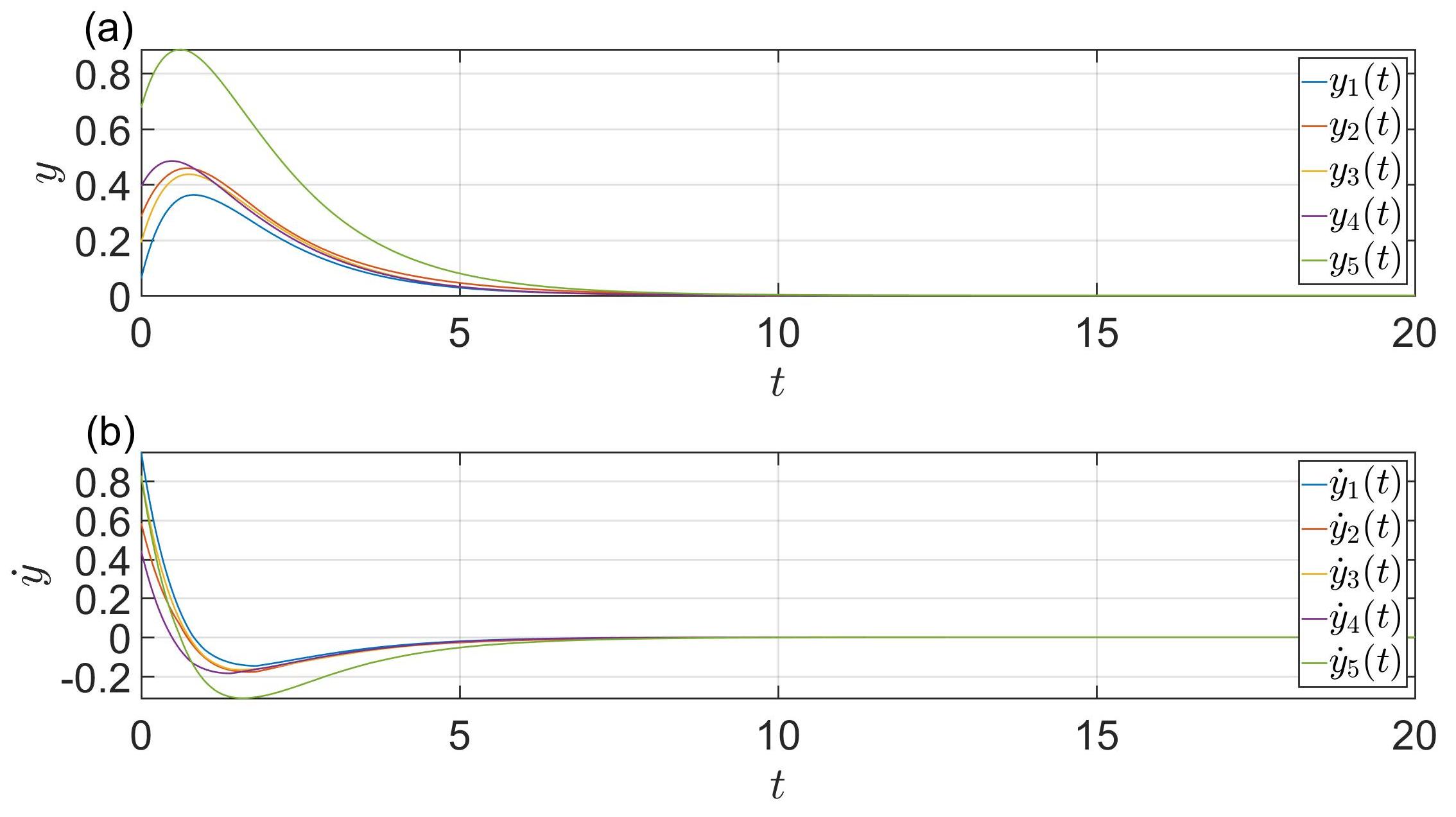}}
\caption{Asymptotic synchronization of neural mass model (NMM) population hybrid network \eqref{netm} of $N=5$ nodes. (a) and (b) dynamics of post-synaptic potentials $y$ and post-synaptic potential derivatives $\dot y$ and of all nodes, respectively. System parameters: $N=5$, $\alpha=1$, $\beta=0.8$, $a=0.5$, $h=2$, $e_0=2.5$, $r=0.56$. Initial conditions are uniformly distributed on the interval $[0;1]$.}
\label{fig2}
\end{figure}

First of all, the network of $N=5$ NMM populations is considered. The system parameters $\alpha$ and $\beta$ are equal to $1$ and $0.8$, respectively. Suppose that all coupling coefficients $a_{ij}$ and all sampling intervals $h_{ij}$ belong to the interval $[0;1]$. Then the network \eqref{netm} can be presented as follows:
\begin{equation} \label{netm2}
\begin{aligned}
\dot y_{i}(t) &= x_{i}(t), \\
\dot x_{i}(t) &= -2\alpha x_{i}(t) -\alpha ^2y_{i}(t)\\
&+\alpha\beta\phi\left[a\sum\limits_{j=1}^N a_{ij}\left(x_{i}(h^*t_{ij}^k)-x_{j}(h^*t_{ij}^k)\right)\right],
\end{aligned}
\end{equation}
where $a>0$ is overall coupling and $h^*>0$ is a sampling interval gain. Consider the graph of the network \eqref{netm} as in Fig.~\eqref{fig1}. This graph is connected and undirected. The number of each vertex corresponds to the number $i$ of each system in the network \eqref{netm}. The number in the rectangle on each edge $(i,j)$ means the value of the corresponding coupling strength $a_{ij}$. The number above each edge $(i,j)$ corresponds to the value of sampling interval $h_{ij}$. The maximum eigenvalue of the corresponding Laplace matrix $\lambda_{\max}$ is equal to $1.6115 a$, while the maximum sampling interval $h$ is equal to $0.9h^*$. Therefore for $a=0.5$ and $h^*=2$ both conditions \eqref{res1}, \eqref{res2} of Theorem~\ref{th2} are fulfilled, i.e. the network \eqref{netm2} is asymptotically synchronized. The results of simulation are presented in Fig.~\ref{fig2}. It can be seen that after some transient approximately equal to $10$ s, the trajectories of system solutions tend to zero which is the equilibrium point. This means that the network is asymptotically synchronized, i.e. the goal \eqref{goal2} is achieved.

\begin{figure}
\center{\includegraphics[width=1\linewidth]{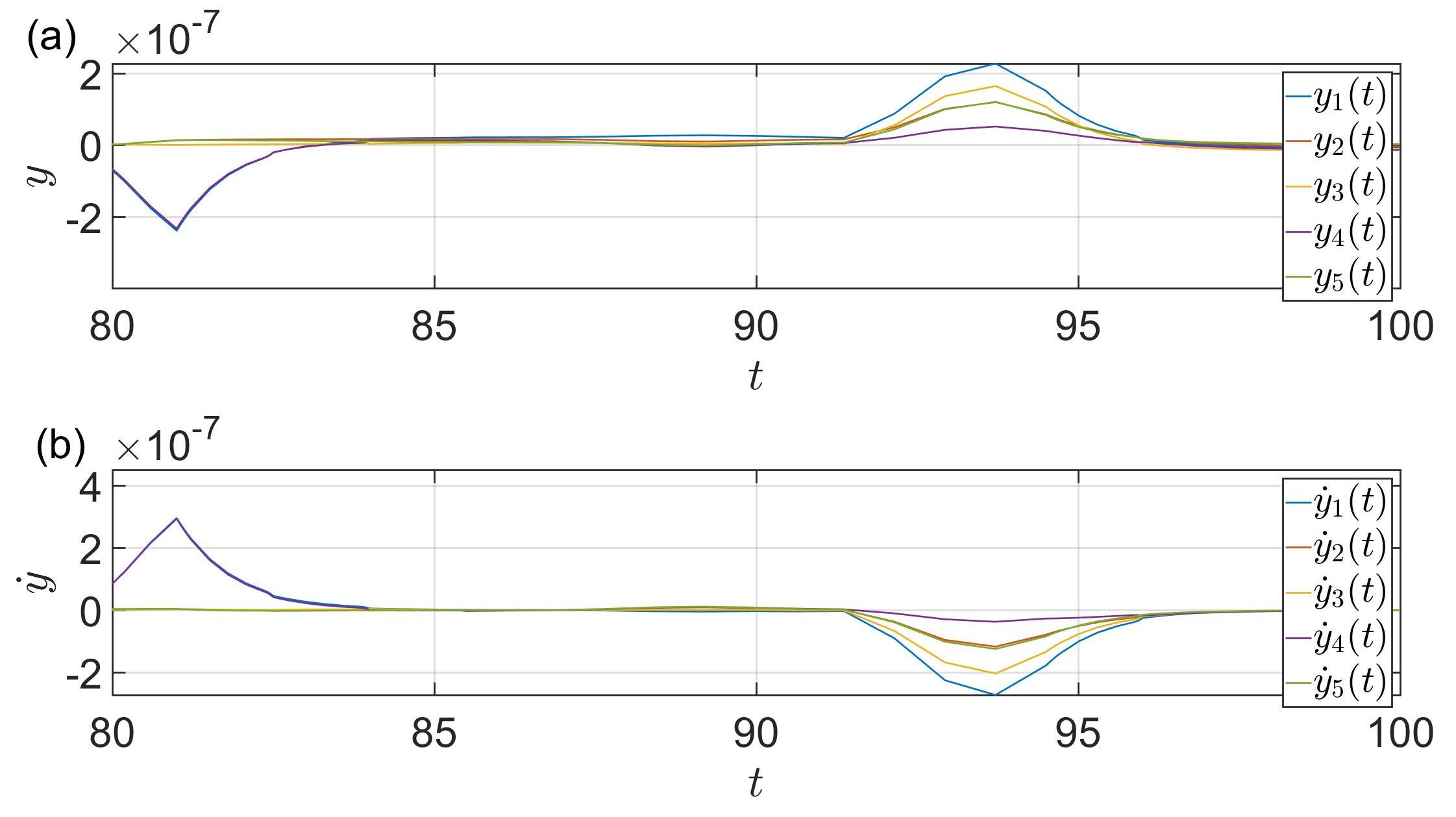}}
\caption{$\varepsilon$-synchronization of neural mass model (NMM) population hybrid network \eqref{netm} of $N=5$ nodes. (a) and (b) dynamics of post-synaptic potentials $y$ and post-synaptic potential derivatives $\dot y$ and of all nodes, respectively. System parameters: $N=5$, $\alpha=1$, $\beta=0.8$, $a=0.5$, $h=15$, $e_0=2.5$, $r=0.56$. Initial conditions are uniformly distributed on the interval $[0;1]$.}
\label{fig3}
\end{figure}

Now choose the same value of overall coupling $a=0.5$, but let the sampling interval gain $h^*$ be equal to $15$. This means that in Theorem~\ref{th2} only inequality \eqref{res1} is feasible, while the inequality \eqref{res2} is not. The results of simulation is given in Fig.~\ref{fig3}. Here the transient is skipped, and one can observe another regime of network \eqref{netm} behavior. The amplitude of oscillations of both state vector components $y_i$ and $\dot y_i$, $i=1,\dots,5$ are bounded by small magnitude $5\times 10^{-7}$. This means that asymptotic synchronization is not reach, however, the current state is called $\varepsilon$-synchronization \cite{FRA07}, i.e. synchronization with some level of precision.

Now choose the same value of overall coupling $a=0.5$, but let the sampling interval gain $h^*$ be equal to $15$. This means that in Theorem~\ref{th2} only inequality \eqref{res1} is feasible, while the inequality \eqref{res2} is not. The results of simulation is given in Fig.~\ref{fig3}. Here the transient is skipped, and one can observe another regime of network \eqref{netm} behavior. The amplitude of oscillations of both state vector components $y_i$ and $\dot y_i$, $i=1,\dots,5$ are bounded by small magnitude $5\times 10^{-7}$. This means that asymptotic synchronization is not reach, however, the current state is called $\varepsilon$-synchronization \cite{FRA07}, i.e. synchronization with some level of precision. Note that we also simulated the dynamics of such network until $t=1000$ s, and we always observe oscillation of the same amplitude.

\begin{figure}
\center{\includegraphics[width=1\linewidth]{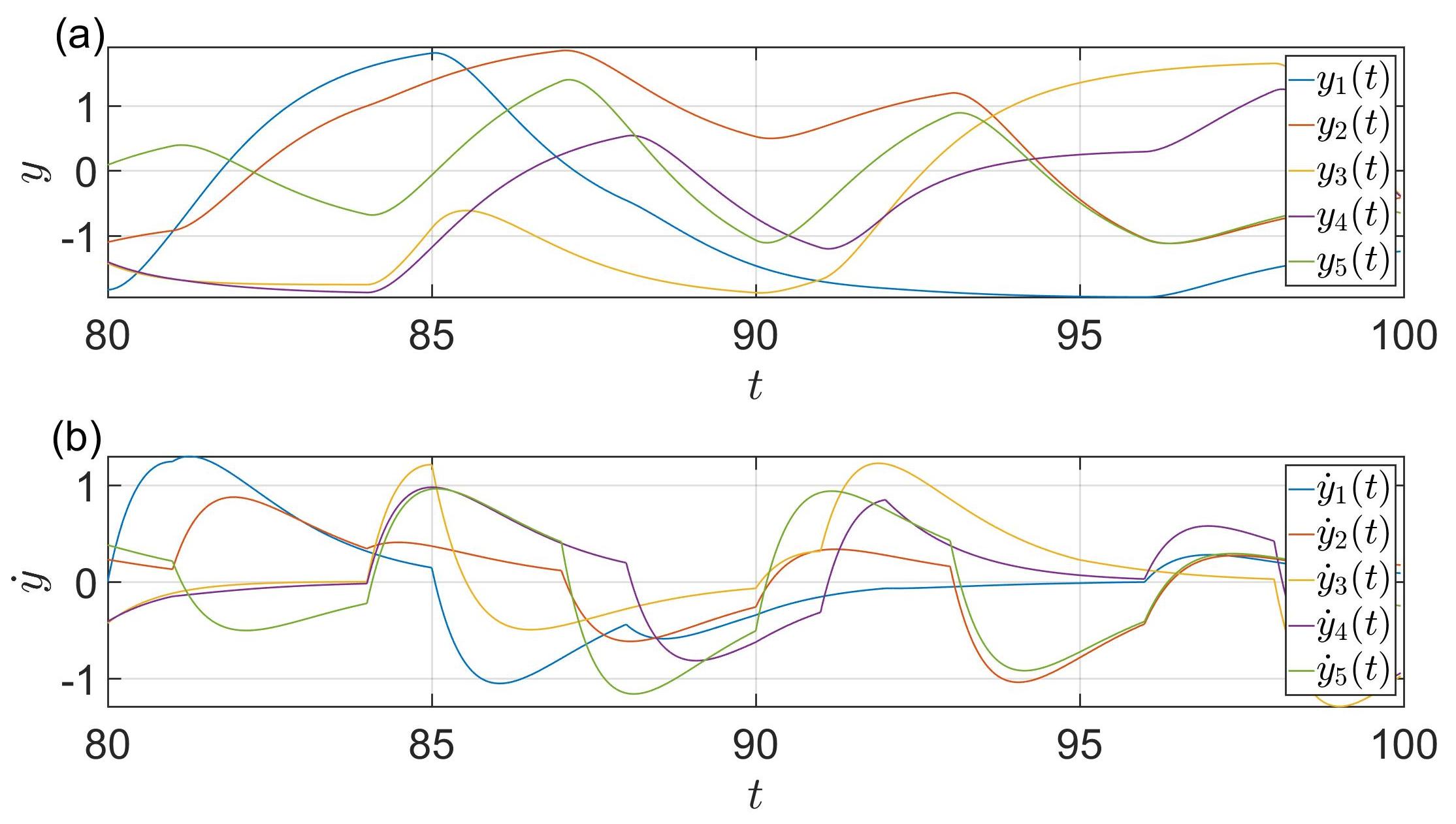}}
\caption{Desynchronization of neural mass model (NMM) population hybrid network \eqref{netm} of $N=5$ nodes. (a) and (b) dynamics of post-synaptic potentials $y$ and post-synaptic potential derivatives $\dot y$ and of all nodes, respectively. System parameters: $N=5$, $\alpha=1$, $\beta=0.8$, $a=20$, $h=10$, $e_0=2.5$, $r=0.56$. Initial conditions are uniformly distributed on the interval $[0;1]$.}
\label{fig4}
\end{figure}

Finally, choose $a=20$ and $h^*=10$. For this case both conditions \eqref{res1}, \eqref{res2} of the Theorem~\ref{th2} are not fulfilled. One can see the results of simulation of such network in Fig.~\ref{fig4}. The transient is also skipped, and there are presented established oscillations of both state vector components $y_i$, $\dot y_i$, $i=1,\dots,5$. This network state is a desynchronization. Note that the uncoupled NMM population is stable, therefore the uncoupled network of identical NMMs is asymptotically synchronized. The presence of couplings with high enough overall coupling may lead to oscillation occurrence. And these oscillations are desynchronized. 

\section{Conclusion}\label{sec:conclusion}

This paper considers the problem of synchronization in networks of neural mass model populations with discrete couplings. The uncoupled neural mass model is continuous systems, therefore the presence of discrete couplings means that the network under consideration is hybrid system. The Mikheev approach \cite{MIK88,FRI92} is used to transform the hybrid system to continuous system with time-varying delay. The sufficient conditions of neural mass model network with delayed couplings were obtained in \cite{PLO22}. Here it has been showed that they are also valid for hybrid neural mass model network, while the corresponding theorem has been proven.

To analyse the hybrid network dynamics the network of $5$ neural mass model populations with discrete couplings has been considered. If the conditions of obtained theorem are fulfilled, namely the maximum eigenvalue of Laplace matrix and maximum sampling interval are small enough, then the network solutions tend to zero equilibrium point, i.e. they are asymptotically synchronized. Increasing of sampling interval leads to the fact the asymptotic synchronization is replaced by $\varepsilon$-synchronization. Increasing of overall coupling, which entails increasing of maximum eigenvalue of Laplacian, leads to oscillation occurrence in the network. The obtained oscillations in the network are desynchronized. Note that obtained theorem gives sufficient conditions of network synchronization, but not necessary. It is expected that these results will be useful for the study of other hybrid networks.

\bibliographystyle{IEEEtran}
\bibliography{IEEEabrv,Plotnikovbib}     

\end{document}